\pdfoutput=1
\documentclass[aps,prb,reprint,showpacs,superscriptaddress,notitlepage,floatfix]{revtex4-1}

\usepackage[pdftex]{graphicx}
\usepackage[pdftex]{epsfig}
\usepackage{epstopdf}
\usepackage{color}
\usepackage{amssymb,amsmath}
\usepackage{graphicx}
\usepackage{dcolumn}
\usepackage{multirow}
\usepackage{hyperref}
\usepackage{siunitx}
\usepackage[dvipsnames]{xcolor}

\usepackage{mathtools}

\usepackage[normalem]{ulem}

\begin{document}

\title{Electronic structure of boron and aluminum $\delta$-doped layers in silicon}
\author{Quinn T. Campbell}
\affiliation{Center for Computing Research, Sandia National Laboratories, Albuquerque NM, USA}
\author{Shashank Misra}
\affiliation{Sandia National Laboratories, Albuquerque NM, USA}
\author{Andrew D. Baczewski}
\affiliation{Center for Computing Research, Sandia National Laboratories, Albuquerque NM, USA}

\begin{abstract}
Recent work on atomic-precision dopant incorporation technologies has led to the creation of both boron and aluminum $\delta$-doped layers in silicon with densities above the solid solubility limit.
We use density functional theory to predict the band structure and effective mass values of such $\delta$ layers, first modeling them as ordered supercells.
Structural relaxation is found to have a significant impact on the impurity band energies and effective masses of the boron layers, but not the aluminum layers.
However, disorder in the $\delta$ layers is found to lead to significant flattening of the bands in both cases.
We calculate the local density of states and doping potential for these $\delta$-doped layers, demonstrating that their influence is highly localized with spatial extents at most 4 nm.
We conclude that acceptor $\delta$-doped layers exhibit different electronic structure features dependent on both the dopant atom and spatial ordering.
This suggests prospects for controlling the electronic properties of these layers if the local details of the incorporation chemistry can be fine tuned.
\end{abstract}

\maketitle

\section{Introduction}
Heavily $\delta$ doping a layer of silicon can be used to create a number of technologically interesting quantum devices for nanoelectronic applications~\cite{oberbeck2002encapsulation,ruess2004toward,qian2006half,ruess2007realization}.
$\delta$ doping creates a narrow, highly doped layer of silicon that generates an approximate $\delta$ potential within the semiconductor device, causing the trapped dopant carriers to form a two-dimensional electron or hole gas. 
Atomic-precision advanced-manufacturing (APAM) techniques combining selective lithography techniques such as scanning tunneling microscopy or pulsed UV laser \cite{katzenmeyer2021photothermal} with surface chemistry and subsequent silicon growth have been used to create highly confined $\delta$-doped layers, typically using phosphorus as the dopant atom~\cite{ruess2004toward,ward2020atomic,bussmann2021atomic}.
Recent work, however, has demonstrated the ability to create $\delta$-doped layers using acceptor precursors, which incorporate aluminum or boron~\cite{vskerevn2020bipolar,radue2021alcl3,dwyer2021b}, raising the intriguing possibility of precision patterning of superconducting regions in silicon~\cite{bustarret2006superconductivity,bourgeois2007superconductivity,shim2014bottom,duvauchelle2015silicon,bonnet2022strongly}. 
The electronic structure of these layers will play a role in determining the performance of any quantum devices utilizing these structures.

Previous work on the electronic structure of $\delta$-doped layers has largely focused on donor-doped layers using a phosphine precursor within the APAM process.
A variety of theoretical methods, from tight binding~\cite{cartoixa2005fermi,lee2011electronic,smith2014electronic}, to planar Wannier orbitals~\cite{qian2005theoretical}, to density functional theory (DFT)~\cite{carter2009electronic,carter2011phosphorus,drumm2013ab1,drumm2013ab2}, to effective mass theory~\cite{drumm2012effective}, have been used to predict the electronic structure of phosphorus $\delta$-doped layers in silicon.
These theoretical predictions have largely been confirmed by subsequent experimental characterization~\cite{miwa2013direct,mazzola2014determining,mazzola2014disentangling,miwa2014valley,mazzola2018simultaneous,holt2020observation}.
The electronic structure of these $\delta$-doped layers has been shown to be crucial to the behavior of quantum devices, with Hagmann \textit{et al.} demonstrating that the electronic structure of Si:P $\delta$-doped layers can be used to measure the confinement of carriers in these layers based on non-destructive electrical transport measurements~\cite{hagmann2018high}.
The success of DFT in describing the electronic structure of phosphorus $\delta$-doped layers suggests that extending this methodology to acceptor $\delta$-doped layers will lead to useful predictions of electronic structure behavior. 

In this manuscript, we use DFT calculations to predict the electronic structure of boron and aluminum $\delta$-doped layers, calculating energies and effective masses of the induced impurity bands.
We demonstrate that boron $\delta$-doped layers induce a significant amount of stress that leads to broadening of the $\delta$ potential and a slight electron-like curvature around the $\Gamma$ point in the highest valence band.
We also calculate the band structure of an acceptor $\delta$-doped layer with significant disorder in the placement of dopant atoms, resulting in significant flattening of the impurity bands. 
We compare these results to the band structure of aluminum $\delta$-doped layers, which induce less stress and come closer to a pure $\delta$ potential and its resulting band structure. 
Our work shows that both the choice of dopant and the spatial ordering of dopants allow for control of the gaps and effective masses of acceptor $\delta$-doped layers, suggesting that these features can be controlled if local incorporation chemistry schemes can be fine tuned.
 
\section{Methods}

All electronic structure calculations are done using the {\sc quantum espresso} package~\cite{giannozzi2009quantum}.
We use norm-conserving pseudopotentials from the PseudoDojo repository~\cite{van2018pseudodojo} and the Perdew-Burke-Ernzerhof exchange-correlation functional~\cite{perdew1996generalized}.
We use kinetic energy cutoffs of 50 Ry and 200 Ry for the plane wave basis sets that define the Kohn-Sham orbitals and charge density, respectively.
We use a 2$\times$2$\times$1 Monkhorst-Pack grid~\cite{monkhorst1976special} to sample the Brillouin zone in our initial self-consistent calculation and then a 4$\times$4$\times$1 Monkhorst-Pack grid for non self-consistent calculations before band structures are calculated.
Unless otherwise noted, we allow the atoms to undergo geometric relaxation at a fixed supercell shape until the interatomic forces fall below a threshold of 50 meV/\AA. 
The cell size is kept fixed to mimic embedding the $\delta$-doped layer in a larger bulk-like silicon crystal.

\begin{figure}
     \includegraphics[width=\columnwidth]{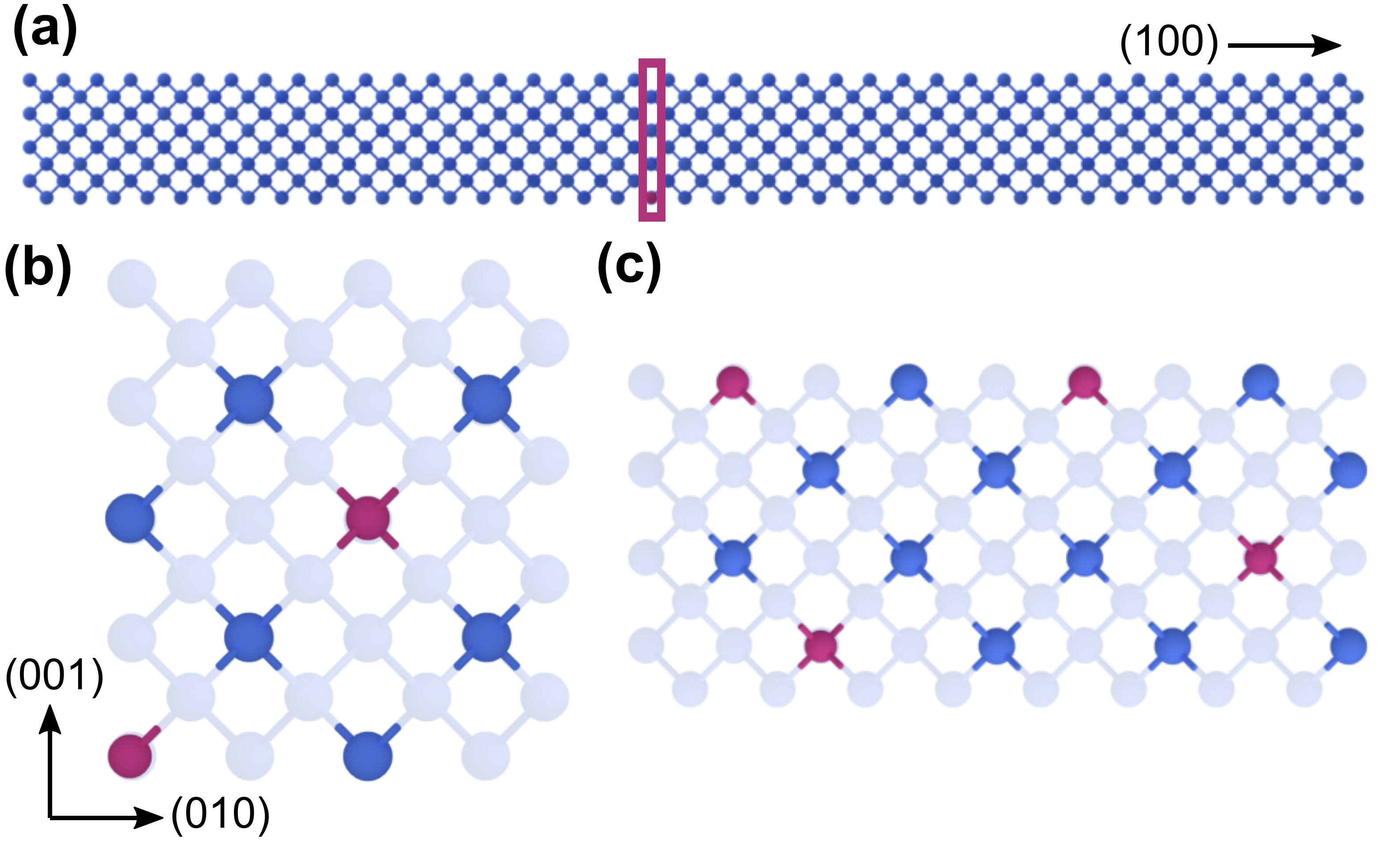}
    \caption{
    (a) Side view of the silicon slab used to calculate the electronic structure of $\delta$-doped layers. The purple box in the middle highlights the location of the $\delta$-doped layer within the overall structure. (b) Top view of the placement of dopant atoms within the ordered 2$\times$2 cell. Silicon atoms are shown in blue, and dopant atoms in purple. The darker spheres indicate the atoms present at the $\delta$ layer and the lighter atoms are silicon at lower layers within the structure. (c) Top view of the placement of dopant atoms within the disordered 4$\times$2 cell. 
    }
    \label{fig:schematic}
\end{figure}

To predict the electronic structure of $\delta$ layers, we create a 2$\times$2 Si(100) slab 110 {\AA} long, with a single layer of 1/4 monolayer concentration of boron or aluminum atoms at substitutional sites, as seen in Fig.~\ref{fig:schematic}a. 
The silicon lattice constant is set at the PBE minimum lattice constant of 5.468 {\AA} based on calculations from the Materials Project~\cite{jain2013commentary}.
Because we are using a plane wave DFT code that enforces periodic boundary conditions in the Kohn-Sham orbitals and potential, the modeled slab repeats in the (100) direction and the $\delta$ layer interacts with an infinite number of its images spaced with a period of 110 {\AA}.
This distance was found to be sufficient to mitigate the effect of these spurious images on our results.
To test the impact of the ordering of the acceptor dopants comprising the $\delta$ layer, two structures are considered: an ordered 2$\times$2 cell shown in Fig.~\ref{fig:schematic}b, and a disordered 2$\times$4 cell shown in Fig.~\ref{fig:schematic}c.
The placement of the atoms in the disordered cell is taken from a Kinetic Monte Carlo (KMC) simulation of dopant placement using a BCl$_3$ precursor within a typical APAM process \cite{campbell2022reaction}.
The same placement of dopant atoms is used for the later aluminum calculations. 

One difficulty of using DFT to predict the $\delta$-doped layer's band structure is disentangling the impact of the placement of dopants in a $\delta$-layer structure from the overall concentration of dopants within the DFT supercell.
We examine this issue in more detail in Appendix A.
We otherwise note that all of our calculations suffer from the well-known tendency of semilocal DFT to underestimate band gaps\cite{perdew1985density}.
For ease of comparison to bulk-like silicon, all of the band diagrams in this manuscript include the band structure of a comparable silicon slab as gray-shaded background.
This is similar to the convention introduced in Ref.~\onlinecite{carter2009electronic} for phosphorus-doped $\delta$ layers.
The band gap we predict for the pure silicon structure is approximately 0.6 eV. 
The conduction band of the systems remains entirely unaffected with the introduction of acceptor layers, and only the valence band is raised. 
Throughout, we reference the band energies to the valence band maximum of silicon to avoid confusion regarding the band gap of the material.
We expect that the conduction band will be unaffected by the acceptor $\delta$ layer and the band gap will increase by $\approx$0.5 eV in real-world materials. 

\section{Results}
\subsection{Boron $\delta$-doped layers}

\begin{figure*}
    \includegraphics[width=\textwidth]{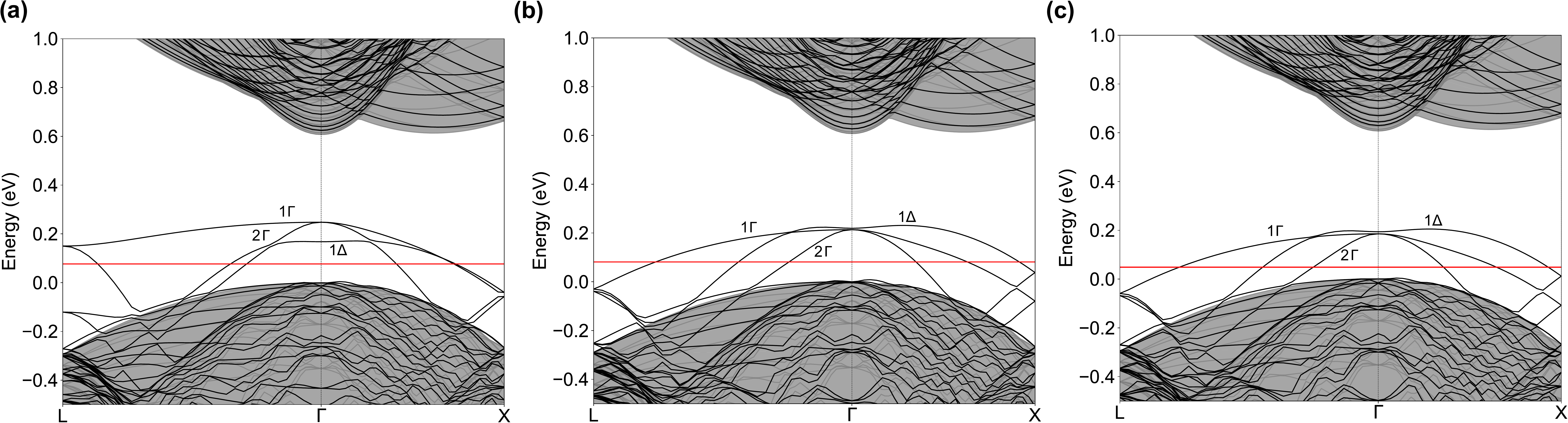}
    \caption{
    The electronic band structure of boron $\delta$-doped layers when (a) all atoms are frozen from their default position in silicon (i.e. no geometry optimization), (b) after the system has been allowed to relax atomic positions in response to stress, and (c) after atomic position relaxation with spin-orbit coupling included in the calculation. In these figures, the gray background shading represents our calculated band structure of bulk silicon and the red line represents the Fermi level. 
    }
    \label{fig:b-bands}
\end{figure*}

The band structure of a $\delta$ layer of boron in silicon without ionic relaxation is shown in Fig.~\ref{fig:b-bands}a.
It is worth noting that we refer all energies to a zero point set by the bulk valence band maximum (VBM) of any given structure.
Furthermore, these band diagrams appear less distinct compared to a typical silicon primitive cell band diagram because the larger supercell used in this work generates Brillioun zone folding. 
There are three impurity bands associated with the $\delta$ layer, two of which form the degenerate valance band maximum (VBM) 248 meV above the bulk-like VBM.
The curvature of the third band (1$\Delta$) becomes slightly positive (electron-like) near the $\Gamma$ point, with the maxima occurring at 0.1$\times$2$\pi$/a in the $X$ direction and a local minima at the $\Gamma$ point. 
Each of the impurity bands is partially occupied, albeit far from the $\Gamma$ point, with the Fermi level at 77 meV above the pure silicon VBM. 

The stress induced in the surrounding silicon by boron $\delta$ layers is significant (148 MPa per unit cell) and thus we examine the impact of relaxing the atomic coordinates on the calculated band structure in Fig.~\ref{fig:b-bands}b.
The lattice constant is held constant, however, to mimic the embedding of the $\delta$ layer within the larger silicon structure. 
Most prominently, the order of the impurity bands away from the VBM is reversed.
Here we determine the character of the bands under the reordering in terms of the two-fold $\Gamma$-point degeneracy (1$\Gamma$ and 2$\Gamma$) and the non-$\Gamma$ VBM (1$\Delta$).
The 1$\Delta$ band now moves up in energy 49 meV to become the VBM of the $\delta$-doped layer electronic structure, this time only 219 meV above the bulk Si VBM, increasing the overall band gap at the $\Gamma$ by 29 meV.
The 1$\Gamma$ and 2$\Gamma$ bands move down by 37 meV, moving to below the 1$\Delta$ band, while still maintaining their double degeneracy.
A small gap of 5 meV is created at the $\Gamma$ point between the 1$\Delta$ and 1$\Gamma$ bands. 
This switching of ordering now means that the top valence band has the opposite curvature as typical around the $\Gamma$ point and that the VBM occurs at 0.14$\times$2$\pi$/a in the $X$ direction. 
The Fermi level for the $\delta$-doped layer nonetheless stays essentially constant at 79 meV.

Spin-orbit coupling has a smaller influence on the band structure of the boron $\delta$-doped layer than the ionic relaxation in response to stress, as shown in Fig.~\ref{fig:b-bands}c where spin-orbit coupling is included in the relaxed structure.
The overall ordering and curvature of bands remains largely constant, with the 1$\Delta$ band remaining as the top impurity band.
The band gap of the $\delta$-doped layer at the $\Gamma$ point increases by 28 meV, with the 1$\Delta$ band now only 200 meV above the bulk VBM at the $\Gamma$ point.
The 1$\Gamma$ and 2$\Gamma$ bands remain doubly degenerate with the gap between them and the 1$\Delta$ point only increasing from 5 to 8 meV, indicating a relatively small contribution of spin-orbit coupling.
The Fermi level again remains constant at 83 meV.
The orbital energies of each band at the $\Gamma$ point are summarized in Table \ref{tab:band-e}.

We also calculate the effective masses at different $k$--points using $m_{ij}^* = \hbar^2 \left(\partial^2 E(k)/\partial k_i \partial k_j \right)^{-1}$, where we approximate the second derivatives using a five-point stencil and the results are reported in Table~\ref{tab:b-eff-mass}.
For the sake of comparison, we note that the heavy- and light-hole effective masses for bulk silicon are respectively 0.49 and 0.16 $m_e$.
We find that the $\Gamma$-point effective masses for the heavy- (1$\Gamma$) and light-hole (2$\Gamma$) impurity bands are respectively lighter and heavier than their bulk counterparts, but only very slightly.
We also note the positive (electron-like) curvature of the 1$\Delta$ band at the $\Gamma$ point.
Most relevant to the transport properties of these layers are the effective masses near the Fermi level at points with $\Delta$ and $\Lambda$ symmetries, where the bands flatten out and have much higher effective masses.

\begin{table}
\centering
\caption{Energy of bands at the $\Gamma$ point for B $\delta$-doped layers, corresponding to the diagrams of Fig.~\ref{fig:b-bands}. All energies are reported in meV relative to the VBM of pure Si.}
\label{tab:band-e}
\begin{tabular}{l c c c }
\hline 
Band & Unrelaxed (2a) & Relaxed (2b) & Relaxed, SOC (2c)\\

\hline 
1$\Gamma$ & 248 & 214 &  186 \\
2$\Gamma$ & 248 & 214 & 186\\
1$\Delta$ & 167 & 219 &  194\\
E$_{\rm F}$ & 77 & 79 &  83\\
E$_{\rm CB}$ & 624 & 627 &  630\\
\hline
\end{tabular}
\end{table}

\begin{table}
\centering
\caption{Effective mass of the different bands at varying $k$ points in the case of a B $\delta$-doped layer which has been calculated including geometry optimization and spin-orbit coupling. All masses are reported in units of m$_e$. The sign convention is taken such that a valence band with negative curvature (i.e. the expected curvature for a valence band) has a positive mass and bands with positive curvatures have negative masses.}
\label{tab:b-eff-mass}
\begin{tabular}{l c c c}
\hline 
Band & $\Gamma$  & E$_{\rm F}$ crossing L   & E$_{\rm F}$ crossing  X \\

\hline 
1$\Gamma$ &  0.43 & 2.07 &  2.65\\
2$\Gamma$ & 0.24 & 3.73 & 0.87\\
1$\Delta$ & --0.56 & 0.95 & 0.88\\
\hline
\end{tabular}
\end{table}

\begin{figure}
    \includegraphics[width=\columnwidth]{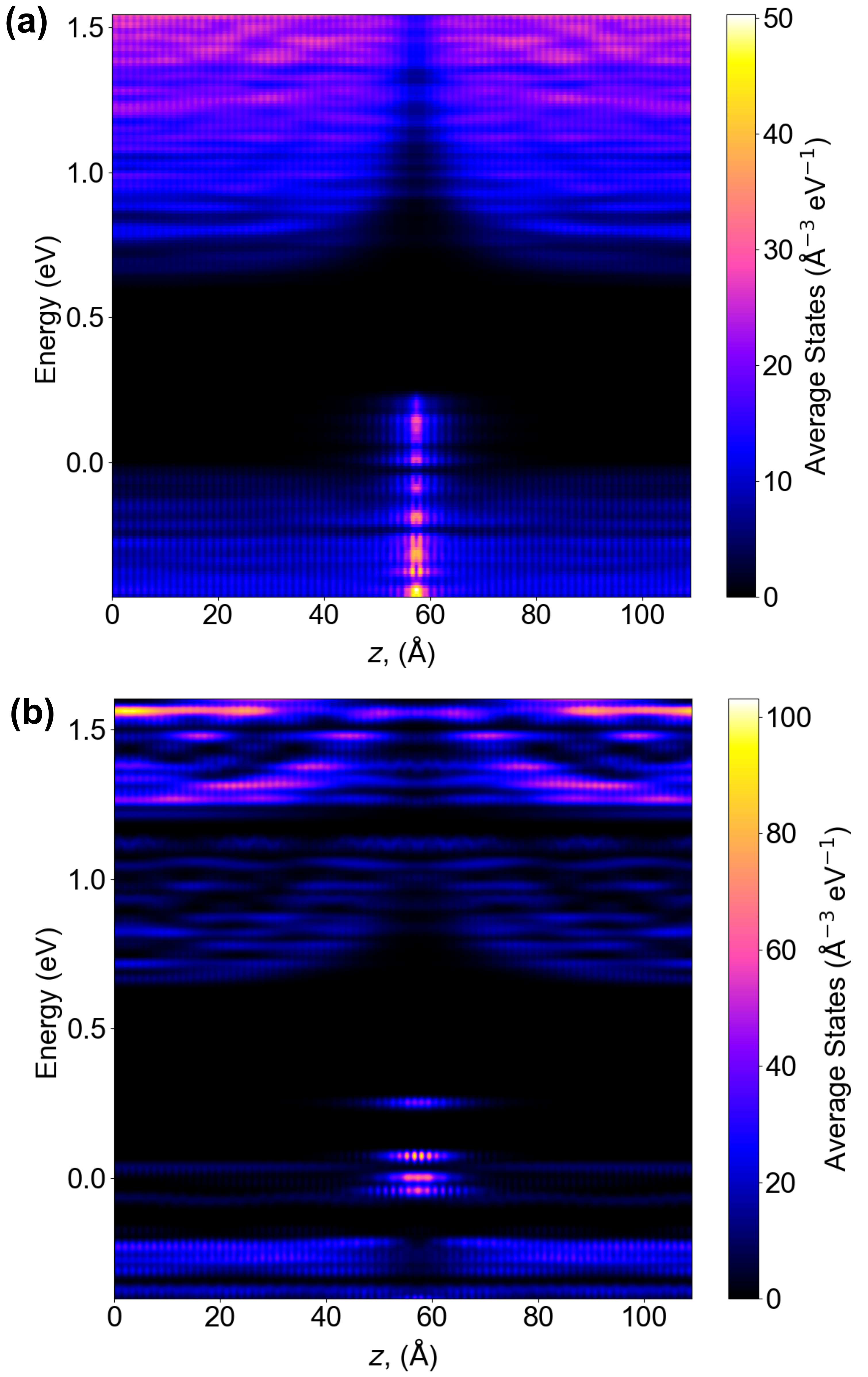}
    \caption{
    Planar average of the local density of states (LDOS) for boron $\delta$-doped layers (a) before and (b) after atomic position relaxation. Dark blue regions indicate no occupied states and red and green regions indicated positions of high occupancy. 
    }
    \label{fig:ldos}
\end{figure}

Next, we consider the spatial extent of the impurity bands by examining the local density of states in Fig.~\ref{fig:ldos}.
Without relaxing the stress created by the $\delta$ layer, the associated states remain well localized within $\approx$1 nm of the layer itself.
In contrast, after relaxation these states delocalize and separate.
While the impurity band edge becomes slightly more diffuse it is largely still concentrated within $\approx$1 nm of the layer itself.
However, the bands below this edge delocalize over the entire $\pm$ 5 nm extent of the supercell and there are evident gaps in energy separating these delocalized states from both the impurity band edge and the bulk valence bands.
The calculated band gap remains a relatively constant 0.6 eV throughout the structure with both the valence and conduction bands forming a $\delta$-potential around the $\delta$-doped layer. 
This band gap corresponds to the typical DFT underestimation of the silicon band gap and could be improved with more computationally expensive methods, e.g. the use of hybrid exchange-correlation functionals~\cite{heyd2003hybrid}.


\begin{figure}
    \includegraphics[width=\columnwidth]{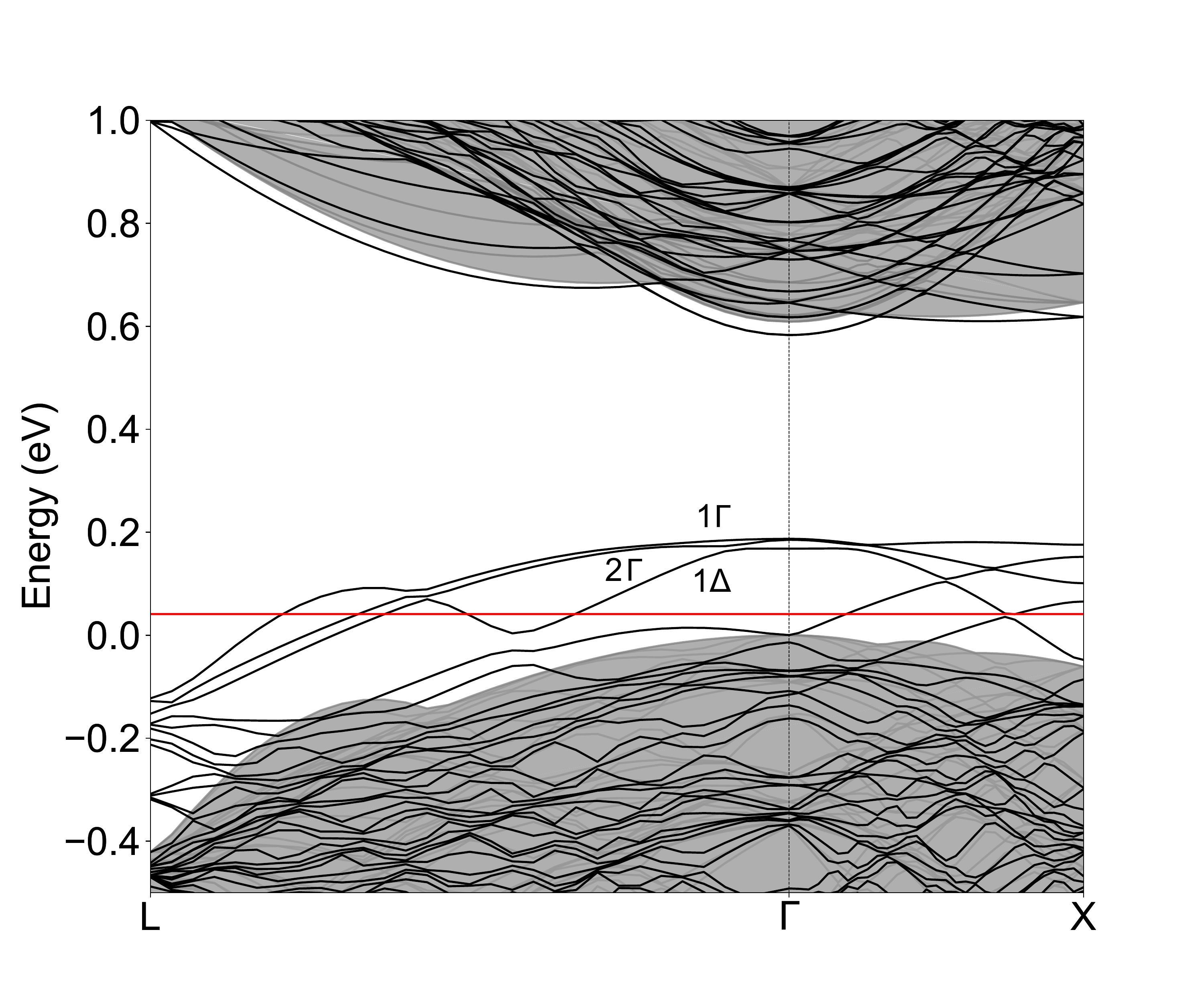}
    \caption{
    The electronic band structure of a disordered boron $\delta$-doped layer with substitutional boron sites taken from a KMC calculation~\cite{campbell2022reaction} and positions subsequently relaxed. The gray background shading is the calculated band structure of bulk silicon and the red line represents the Fermi level.
    }
    \label{fig:disordered}
\end{figure}

While we have seen that structural relaxation tends to encourage delocalization of the impurity bands, it is worth noting that these results have assumed an ordered supercell.
We next consider the impact of realistic disorder on our results by studying the KMC-derived layer displayed in Fig.~\ref{fig:schematic}c, but allowing the structure to relax.
We do not include spin-orbit coupling in these calculations. 
The associated band structure shown in Fig.~\ref{fig:disordered}.
In contrast to the ordered structures shown before, the bands are flattened.
The overall band gap increases by 31 meV compared to the relaxed ordered case (equivalently without spin-orbit coupling included).
The 1$\Delta$ band is again the bottom impurity band, meaning the VBM does occur at the $\Gamma$ point in this structure. 
The disordered structure also lifts the degeneracy of all the bands, creating a gap of 3 meV between the 1$\Gamma$ and 2$\Gamma$ band and a gap of 17 meV between the 2$\Gamma$ and 1$\Delta$ bands.
The Fermi level is also notably pulled down by 41 meV, placing it only slightly above the bulk valence band maximum. 
The bands are generally much flatter and have higher effective masses, although the 2$\Gamma$ band still exhibits significant curvature around the $\Gamma$ point, as detailed in Table \ref{tab:b-disordered-e}.
The 1$\Delta$ band continues to exhibit opposite curvature as to be expected around the $\Gamma$ point, resulting in a large negative effective mass of --2.32.
Such large effective masses might be useful for realizing high kinetic inductance superconducting elements~\cite{masluk2012microwave,niepce2019high}.

\begin{table}
\centering
\caption{The energy and effective mass of bands in a disordered B $\delta$-doped layer structure. Energies are given in units of meV relative to the Si VBM, and effective masses are given in units of m$_e$. The sign convention is such that a valence band with negative curvature (i.e. the expected curvature for a valence band) has a positive mass and bands with positive curvatures have negative masses.}
\label{tab:b-disordered-e}
\begin{tabular}{l c c }
\hline 
Band & Energy at $\Gamma$ & Effective mass at $\Gamma$ \\
\hline 
1$\Gamma$ & 188  & 1.46 \\
2$\Gamma$ & 185 & 0.36 \\
1$\Delta$ & 168 & --2.32 \\
E$_{\rm F}$ & 42 &  \\
E$_{\rm CB}$ & 583 &   \\
\hline
\end{tabular}
\end{table}

%
%
\subsection{Aluminum $\delta$-doped layers}

\begin{figure}
    \includegraphics[width=\columnwidth]{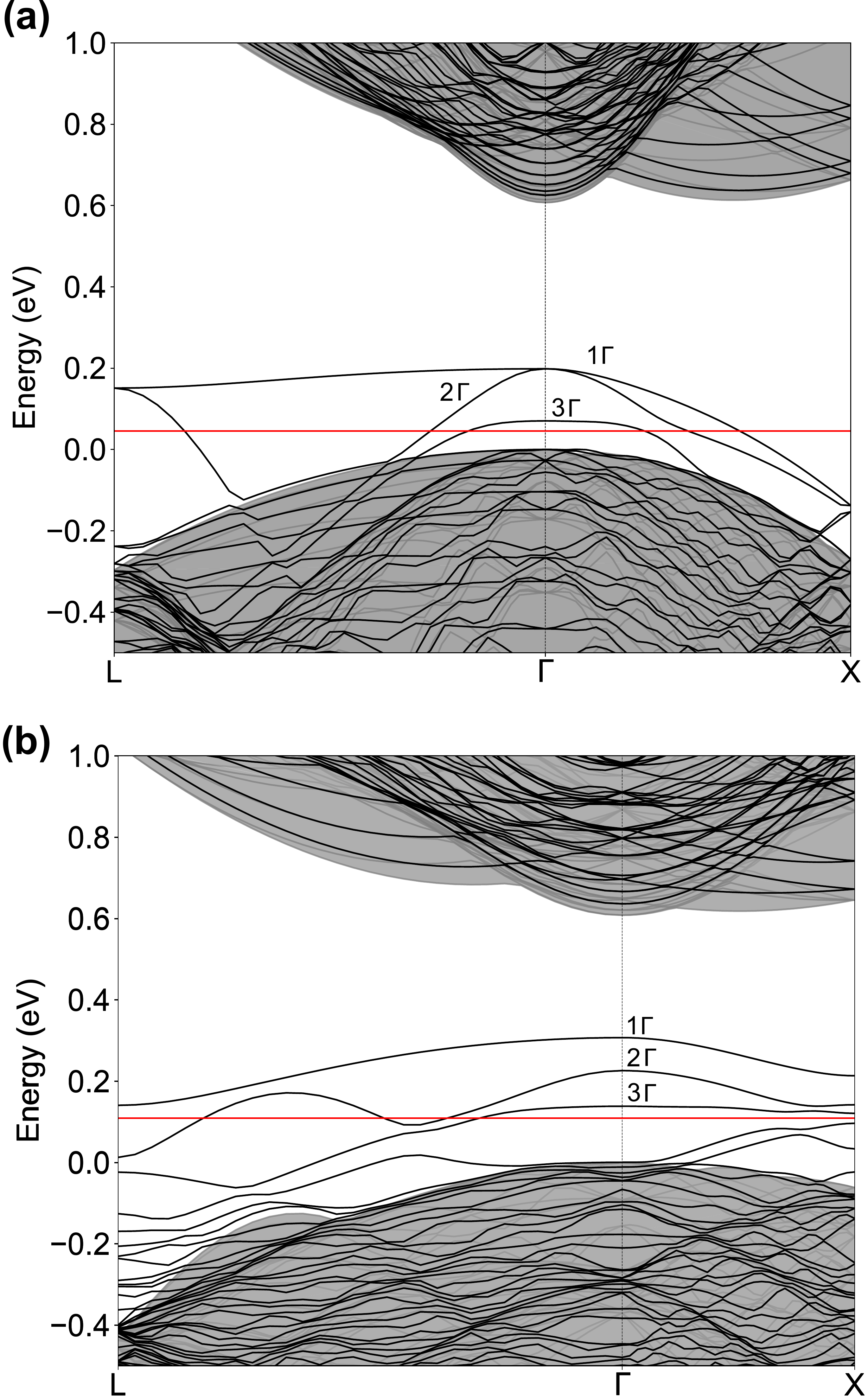}
    \caption{
    The electronic structure after atomic relaxation of an aluminum $\delta$-doped layer that is (a) ordered and (b) disordered. The gray background shading represents our calculated band structure of bulk silicon and the red line represents the Fermi level. 
    }
    \label{fig:Al}
\end{figure}

Aluminum $\delta$-doped layers experience significantly less distortion due to stress than boron $\delta$-doped layers. 
The calculated band structure for an aluminum $\delta$-doped layer, which has undergone geometric optimization and includes spin-orbit coupling, is shown in Fig.~\ref{fig:Al}. 
This is notably similar to the band structure of the boron $\delta$-doped layer before any stress induced relaxation is included, suggesting that an aluminum $\delta$-doped layer will have a much smaller impact on its neighboring layers (and indeed induces less stress, at 85 MPa per unit cell).
This is consistent with chemical intuition-- an aluminum atom is closer in size to a silicon atom than a boron atom and therefore generates less strain to the lattice.
Even with spin-orbit coupling included, there is still a degeneracy in the topmost 1$\Gamma$ and 2$\Gamma$ bands. 
The bands are also pulled closer to the bulk valence bands with the band gap of the structure increasing by 28 meV and the Fermi level being pulled down by 33 meV. 
The exact values of the band and Fermi level energies are displayed in Table \ref{tab:al-energy-eff-mass} along with the effective masses.
The top 1$\Gamma$ and bottom 3$\Gamma$ band are both extremely flat with effective masses $\gtrapprox$ 2 m$_e$.
The middle band has an effective mass of 0.32 m$_e$, between the heavy- and light-hole masses in bulk silicon.

\begin{table}
\centering
\caption{The energy and effective mass of bands in an ordered Al $\delta$-doped layer structure. Energies are given in units of meV relative to the VBM, and effective masses are given in m$_e$.}
\label{tab:al-energy-eff-mass}
\begin{tabular}{l c c }
\hline 
Band & Energy at $\Gamma$ & Effective mass at $\Gamma$ \\

\hline 
1$\Gamma$ &  199 & 1.96 \\
2$\Gamma$ & 199 & 0.32 \\
3$\Gamma$ & 71 & 2.68 \\
E$_{\rm F}$ & 46 &  \\
E$_{\rm CB}$ & 625 &   \\
\hline
\end{tabular}
\end{table}

\begin{table}
\centering
\caption{The energy and effective mass of bands in a disordered Al $\delta$-doped layer structure. Energies are given in units of meV, and effective masses are given in m$_e$. }
\label{tab:al-disorder-energy-eff-mass}
\begin{tabular}{l c c }
\hline 
Band & Energy at $\Gamma$ & Effective mass at $\Gamma$ \\

\hline 
1$\Gamma$ & 307 & 1.75 \\
2$\Gamma$ & 226 & 0.36 \\
3$\Gamma$ & 138 &  5.33 \\
E$_{\rm F}$ & 109 &  \\
E$_{\rm CB}$ & 636 &   \\
\hline
\end{tabular}
\end{table}

Disordered aluminum $\delta$-doped layers behave qualitatively similarly to disordered boron $\delta$-doped layers, with the induced bands both being flattened significantly, as demonstrated in Fig.~\ref{fig:Al}b.
However, the valence bands are significantly raised and the overall band gap is lowered by 108 meV in comparison with the ordered aluminum structure.
All degeneracies in the bands at the $\Gamma$ point are lifted with the 1$\Gamma$ and 2$\Gamma$ bands separated by 81 meV, and the 2$\Gamma$ and 3$\Gamma$ bands separated by 88 meV. 
The Fermi level in the case now rests below the 3$\Gamma$ point, introducing another band above the Fermi level.
The 1$\Gamma$ and 3$\Gamma$ bands flatten significantly with effective masses over 1 m$_e$ for both, with the exact values given in Table \ref{tab:al-disorder-energy-eff-mass}.
The 2$\Gamma$ band remains highly curved, with an effective mass near the same magnitude as the 
This flattening of the 1$\Gamma$ and 3$\Gamma$ bands is also potentially promising for superconducting applications requiring large kinetic inductances~\cite{masluk2012microwave,niepce2019high}.

\begin{figure}
    \includegraphics[width=\columnwidth]{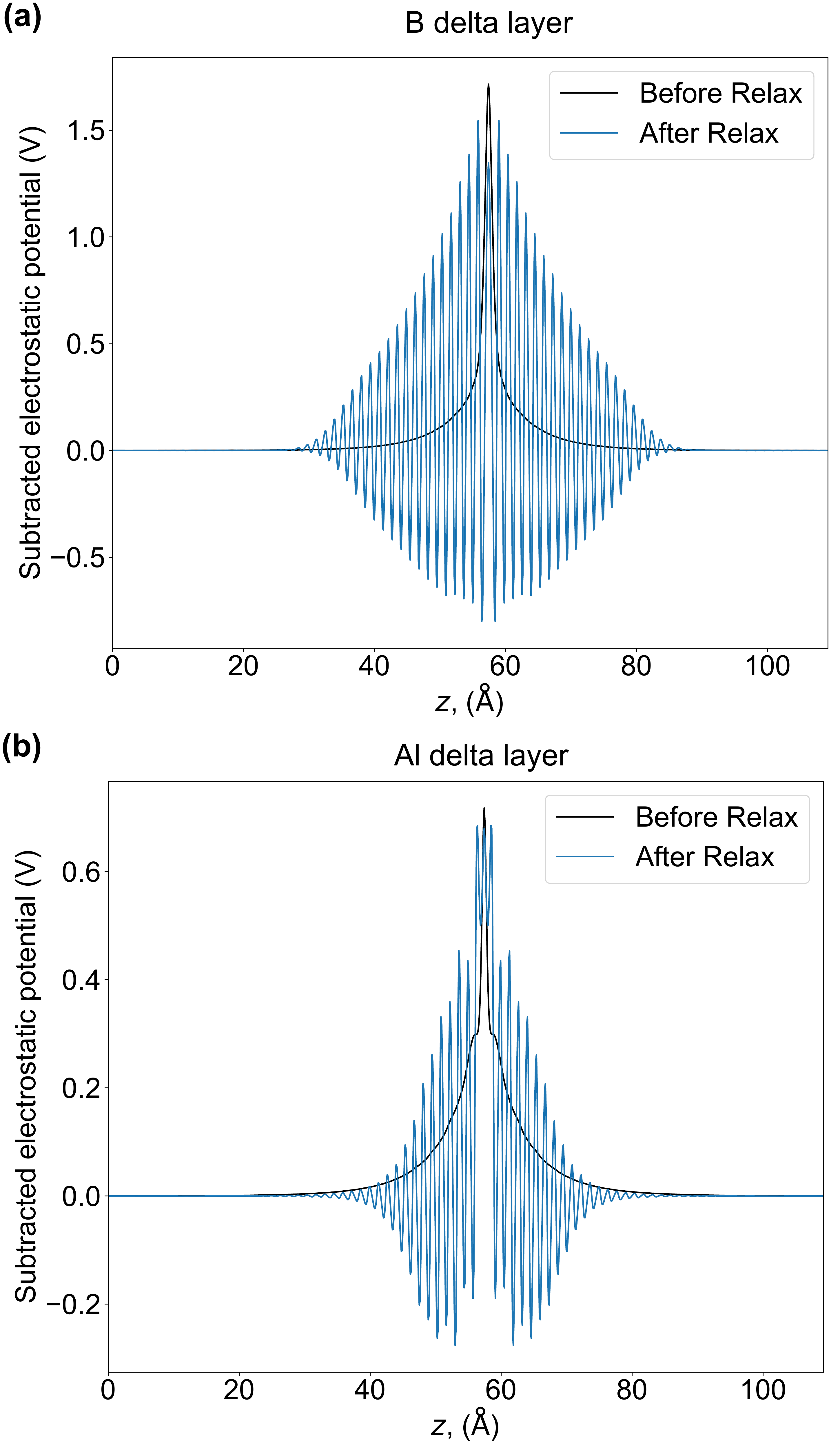}
    \caption{
    The doping potential of ordered (a) boron and (b) aluminum $\delta$-doped layers both before and after relaxation, causing atomic displacement. The aluminum doping potential much more closely resembles a pure $\delta$ potential within silicon.
    }
    \label{fig:potential}
\end{figure}

The difference in the extent of the impact of geometry optimization on boron versus aluminum $\delta$-doped layers can be most clearly seen in the doping potential for each, shown in Fig.~\ref{fig:potential}.
To calculate the doping potential we subtract the electrostatic potential of the  $\delta$-doped layer after geometry optimization from the electrostatic potential of a pure silicon slab.
The geometry optimization in both the aluminum and boron $\delta$ layers leads to widespread displacement in the exact position of neighboring silicon atoms which accounts for the rapid oscillations in doping potential after relaxation seen in Fig.~\ref{fig:potential}.
These displacements are much more pronounced around the boron $\delta$-doped layer, however; the aluminum doping potential has a clear peak in the middle corresponding to the $\delta$-doped layer.
The boron doping potential, in contrast, extends $\sim$ 10 nm further out from the $\delta$ layer in either direction and with a much greater magnitude of potential at these tails of influence. 
The smaller impact of atomic displacement on aluminum systems matches previous work on phosphorus $\delta$-doped layers, where DFT work has shown that relaxation of the underlying position of the donors leads to negligible changes in the electronic structure.\cite{drumm2013ab1}
This further emphasizes the unique nature of boron as a $\delta$-doped layer dopant: it induces significant atomic displacement in the surrounding system that leads to widespread changes in the charge, position, and potential of neighboring layers which in turn causes a significant shift in the overall band structure, a feature which has not been seen in any other $\delta$-doped layer dopants.
Aluminum does not induce anywhere near the same amount of displacement and thus the corresponding $\delta$-doped layer behaves much more like a pure $\delta$ potential in silicon.

Overall, while atoomic displacement plays little role in the band structure of aluminum $\delta$ layers, the exact placement of the dopant atoms within the $\delta$ layer will have a large impact on the curvature, and thus effective mass, of the resulting electronic structure.
We further explore the band structure of an Al `line' structure~\cite{radue2021alcl3} within the $\delta$-doped layer in Appendix B.

\section{Conclusion}

We have reported the results of DFT calculations of the band structure of boron and aluminum $\delta$-doped layers embedded in silicon.
They demonstrate that the band structure of boron $\delta$-doped layers depends strongly on the detailed geometry of the $\delta$ layer, with both geometry optimization in response to induced stress and disorder in the location of substituted boron producing effects. 
The band structure calculations for aluminum $\delta$-doped layers show that these layers induce less atomic displacement due to stress, and consequently behave much more like a pure $\delta$ potential in silicon.

It should be noted that this DFT model will not correspond perfectly to the behavior of real-world $\delta$-doped systems. 
Our silicon does not include any background level of defects/dopants as this would be computationally prohibitive within our model. 
Furthermore, Mazzola \textit{et al.} have recently shown that non--perfect confinement of donors to a single layer can lead to less valley splitting and additional bands crossing the Fermi level beyond what DFT predicts.\cite{mazzola2020sub}
Additionally, we do not include any chlorine that may also become trapped inside the silicon during a $\delta$-doping process which uses aluminum or boron trichloride as dopant precursors.
Significant levels of chlorine may significantly impact the electronic characteristics of these layers.

Our work indicates that different acceptor chemistries can be used to alter the electronic structure of $\delta$-doped layers.
While boron $\delta$-doped layers avoid degeneracy at the top valence band (particularly in disordered states) and move close to a direct band gap, they also induce significant stress within the system that can lead to inversion of the curvature of the top valence band.
In contrast, ordered aluminum $\delta$-doped layers are expected to be more similar to a pure $\delta$ potential, and might provide greater control over factors like $\delta$ potential width. 
They do, however, introduce degeneracies at the top levels of the induced valence bands and have relatively large effective masses.
Disordered aluminum $\delta$-doped layers, in contrast, have no degeneracy, but even larger effective masses.
Recent work has proposed using a guided self-assembly process to create atomically precise patterning of $\delta$-doped layers,\cite{radue2022dopant} which would enable the creation of the ordered band structures similar to those displayed here, provides another axis for controlling the electronic structure.
Given this choice of both ordered versus disordered and boron versus aluminum, we conclude that the band structure of an acceptor based $\delta$-doped layer can be designed for desirable characteristics such as absence of degeneracy at the $\Gamma$ point or large/small effective mass.

\begin{acknowledgments}
We gratefully acknowledge useful conversations with
Evan Anderson,
Ezra Bussmann,
Jeff Ivie,
Tzu-Ming Lu,
Jonathan Moussa, 
Scott Schmucker,
and Bob Butera. 
This work was supported by the Laboratory Directed Research and Development program at Sandia National Laboratories under project 213017 (FAIR DEAL) and project 226347.
This article has been authored by an employee of National Technology \& Engineering Solutions of Sandia, LLC under Contract No. DE-NA0003525 with the U.S. Department of Energy (DOE).
The employee owns all right, title and interest in and to the article and is solely responsible for its contents. 
The United States Government retains and the publisher, by accepting the article for publication, acknowledges that the United States Government retains a non-exclusive, paid-up, irrevocable, world-wide license to publish or reproduce the published form of this article or allow others to do so, for United States Government purposes. 
The DOE will provide public access to these results of federally sponsored research in accordance with the DOE Public Access Plan https://www.energy.gov/downloads/doe-public-access-plan .

\end{acknowledgments}

\bibliography{references}

\clearpage
\widetext
\appendix

\section{Concentration Effects within DFT}
\label{app:conc-effects}
\begin{figure}
    \includegraphics[width=0.5\columnwidth]{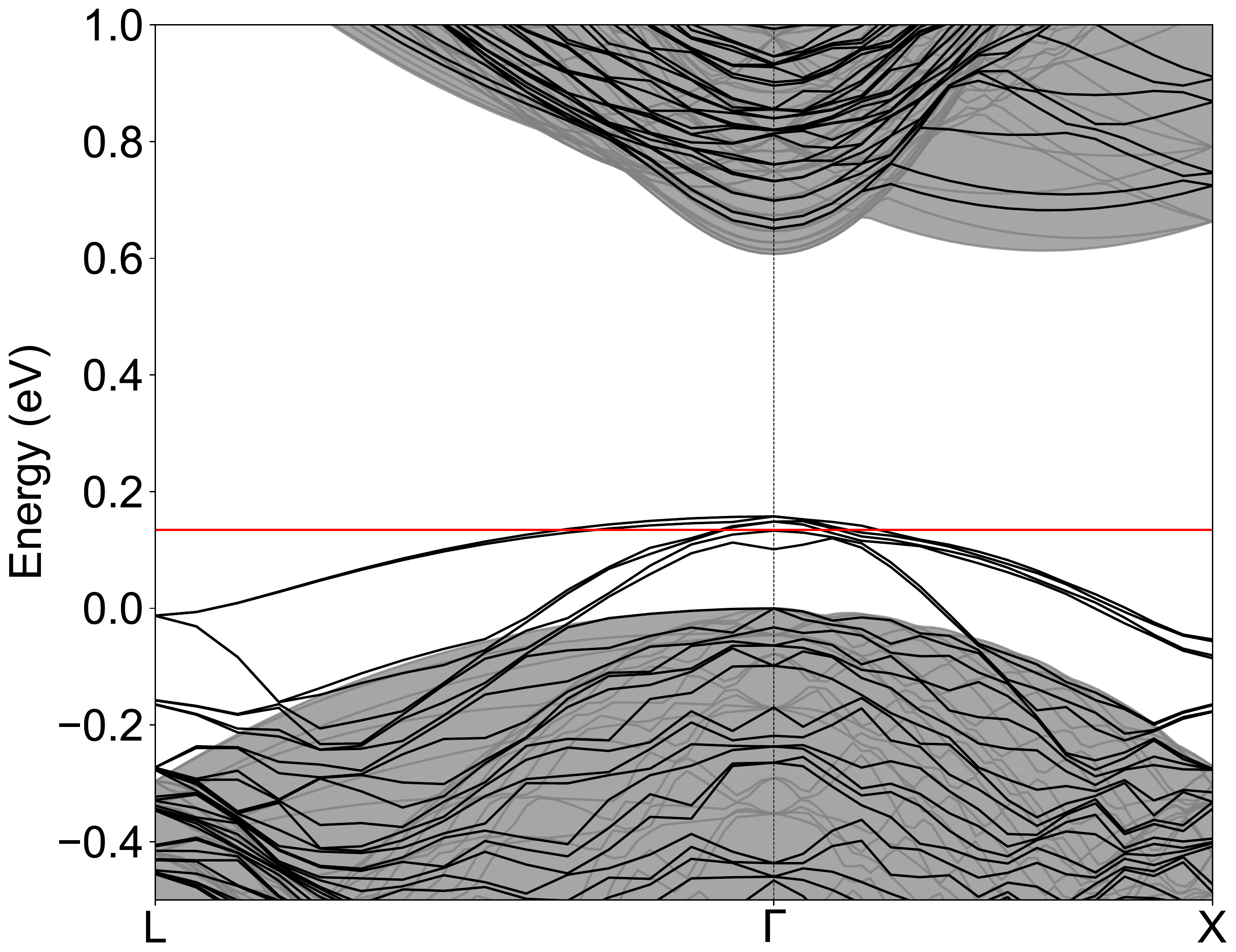}
    \caption{ 
    The band structure of a system with the same supercell as used throughout the work and the same concentration of dopants, but the dopants are scattered through the system instead of fixed in a $\delta$-doped layer.
    }
    \label{fig:not-delta}
\end{figure}

One factor complicating the use of DFT to predict the band structure of $\delta$-doped layers is disentangling the effect of the arrangement of dopants in a $\delta$ layer from the concentration effect of a finite number of dopants in a periodically repeated cell.
For the supercell illustrated in Fig.~\ref{fig:schematic}a and b, the concentration of dopants within the $\delta$-doped layer is 1/4 ML.
Within the entire cell, however, the concentration is \SI{7.64e19}{cm^{-3}}, a relatively high concentration for a typical semiconductor.
It might be expected that for this high level of concentration, we should see large induced valence bands, regardless of whether the boron atoms are forming a $\delta$-doped layer or not.
To test this effect, in Fig.~\ref{fig:not-delta}, we predict the band structure of the 2$\times$2 supercell used throughout the work with the same concentration of B dopants (Figs.~\ref{fig:schematic}a and b), but with the positions randomly determined, instead of forming a controlled $\delta$-doped layer.

The shape of the impurity bands is very similar to the predicted shape of a boron $\delta$-doped layer (see Fig.~\ref{fig:b-bands}a), with a doubling of the bands similar to what Carter \textit{et al.} report for two interacting phosphorus $\delta$-doped layers \cite{carter2013electronic} (indeed, given the setup of our structure, this could alternately be viewed as the band structure of two spatially separated $\delta$-doped layers with a 1/8 ML concentration of acceptors).
The magnitude of the impurity bands is somewhat reduced with a 1$\Gamma$ energy of 158 meV, compared to a 1$\Gamma$ energy of 248 meV in the $\delta$-doped structure.
This demonstrates that while a significant amount of the band structure can be attributed to the concentration effect of using a limited supercell within DFT, the magnitude and clarity of the structure is still related to the arrangement of atoms into a $\delta$-doped layer.
We expect a similar effect would be seen for an analogously randomized Al structure. 
Our current technique for measuring the $\delta$-doped layer electronic structure matches that of the literature with phosphorus $\delta$-doped layers \cite{carter2009electronic,carter2011phosphorus,drumm2013ab1,drumm2013ab2}.
These theoretical techniques have largely been subsequentally confirmed by experimental characterization \cite{miwa2013direct,mazzola2014determining,mazzola2014disentangling,miwa2014valley,mazzola2018simultaneous,holt2020observation}.
We thus have some confidence that while the impact of $\delta$-doped layers and the concentration of dopants in a DFT supercell are unavoidably somewhat entangled, our work nevertheless provides a useful first-order approximation of the electronic structure of boron and aluminum $\delta$-doped layers.

\section{Line features }
\label{app:line_features}
\begin{figure}
    \includegraphics[width=\columnwidth]{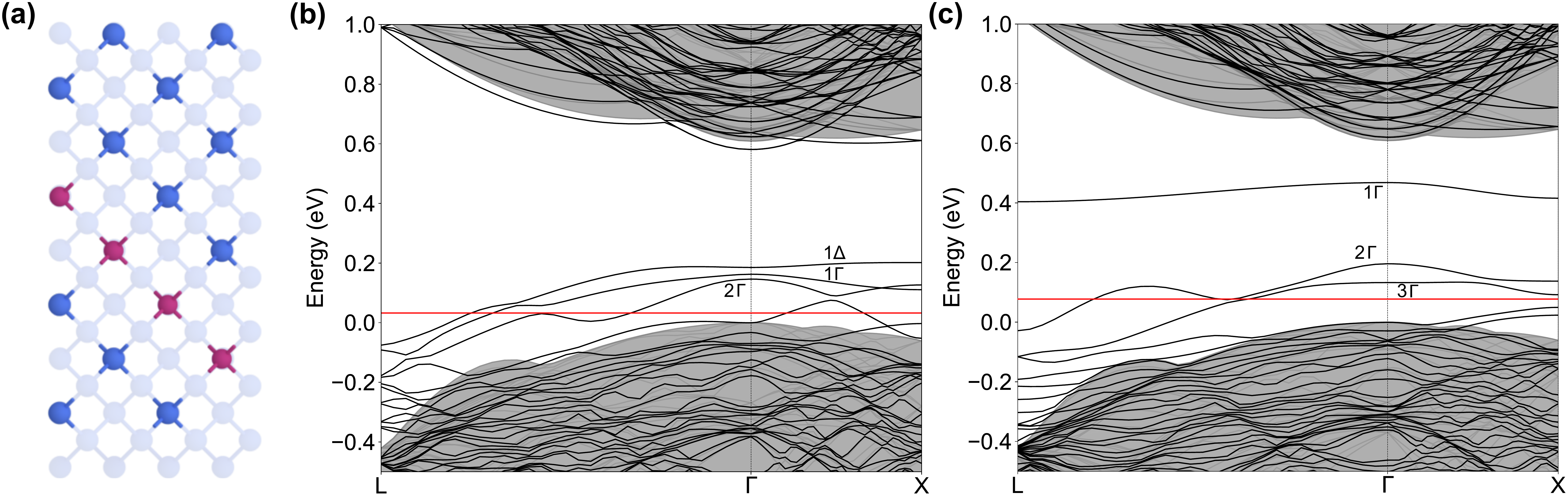}
    \caption{ 
    (a) Top view of the placement of dopant atoms making a line feature within a 2$\times$4 cell. (b) Band structure of boron $\delta$-doped line feature. (c)  Band structure of aluminum $\delta$-doped line feature.
    }
    \label{fig:line-feature}
\end{figure}

Radue \textit{et al.} noted that using AlCl$_3$ as an APAM precursor leads to the formation of large ``line'' structures, where AlCl fragments migrate to form a linear superstructure on the silicon surface~\cite{radue2021alcl3}.
We accordingly examine the band structure of $\delta$-doped layers with a linear structure of dopant atoms for both boron and aluminum, as shown in Fig.~\ref{fig:line-feature}.
While these line structures have not been observed in boron $\delta$-doped layers, they nonetheless provide a useful comparison to the observed aluminum structures.

The band structure of the boron line structure, shown in Fig.~\ref{fig:line-feature}a, combines the features of the relaxed ordered and disordered band structures shown earlier.
Similar to the fully relaxed, ordered structure, the 1$\Delta$ band is pulled to be the highest energy valence band, creating an inversion of the curvature around the $\Gamma$ point, and moving the VBM to 0.12 $\times$2$\pi$/a in the $X$ direction.
Similar to the disordered structure, however, the impurity bands exhibit no degeneracy with a separation of 23 meV between the 1$\Delta$ and 1$\Gamma$ band and 17 meV between the 1$\Gamma$ and 2$\Gamma$ band.
While the 1$\Delta$ band is highly flat with an effective mass of --1.64, the 1$\Gamma$ and 2$\Gamma$ bands are curved with effective masses of 0.51 and 0.61, respectively.
The exact values of the bands and effective masses are displayed in Table ~\ref{tab:b-line-energy-eff-mass}.
Overall, our results highlight that the exact band structure of boron $\delta$ layers is highly dependent on the exact placement of boron atoms.

In aluminum line structures, however, the top impurity band is moved dramatically up, as shown in Fig.~\ref{fig:line-feature}b. 
The 1$\Gamma$ band is moved to 468 meV above the bulk VBM, with a separation of 272 meV between it and the next 2$\Gamma$ band. 
Both the 1$\Gamma$ and 3$\Gamma$ bands are fairly flat, with effective masses greater than unity.
The 2$\Gamma$ band is curved, however, with an effective mass of only 0.4. 
This indicates that the aluminum line feature, which is favored in experiments, has dramatic results on the accompanying electronic structure.
The exact values for the impurity band energies and effective masses can be found in Table ~\ref{tab:al-line-energy-eff-mass}.
It should again be noted that these band structures were calculated at the GGA level and future work using more accurate approximations such as hybrid functionals would be useful to accurately determine the exact shift of the impurity bands, particularly relative to the larger band gap. 
We should also note that due to the periodic nature of these calculations, the line feature of four aluminum atoms is repeated in every supercell, but does not form a continuous line. 
Radue \textit{et al.} found that these line structures were typically only three aluminum atoms long~\cite{radue2021alcl3}, so the proposed line structure of Fig.~\ref{fig:schematic}d used throughout this work will likely be somewhat close to the experimentally realized structure. 

While this line leads to an unusual band structure when using aluminum acceeptors, our finding of an elevated 1$\Gamma$ band remains consistent even when a variety of methodologies are used.
We have examined the same line feature using rotated supercells as well as supercells extended in the further in the slab direction, increasing the distance between periodic images and still found either the same elevated band or a band spanning the entire width of the band gap.
To investigate whether this was due to unusual ordering of the atomic positions, we additionally ran ab-initio molecular dynamics (AIMD) at a temperature of 25 C for 0.5 ps and then measured the band structure, producing the same elevated band.
Finally, we also used SCAN pseudopotentials~\cite{yao2017plane}, which have been shown to improve the band gap estimation, which similarly produced a 1$\Gamma$ band which spanned the entire band gap.
While it seems likely that real world devices with this line feature will have a band structure more closely resembling those seen throughout the main text of this work, the multiplicity of methods giving the same results indicate that some interesting physics is at work in this case and further work is warranted.

\begin{table}
\centering
\caption{The energy and effective mass of bands in a Boron $\delta$-doped layer structure forming a line feature. Energies are given in units of meV, and effective masses are given in m$_e$. }
\label{tab:b-line-energy-eff-mass}
\begin{tabular}{l c c}
\hline 
Band & Energy at $\Gamma$ & Effective mass at $\Gamma$  \\

\hline 
1$\Delta$ & 186  &  --1.64\\
1$\Gamma$ & 163 & 0.61 \\
2$\Gamma$ & 146 & 0.51 \\
E$_{\rm F}$ & 33 & \\
E$_{\rm CB}$ & 581 &  \\
\hline
\end{tabular}
\end{table}

\begin{table}
\centering
\caption{The energy and effective mass of bands in an Aluminum $\delta$-doped layer structure forming a line feature. Energies are given in units of meV, and effective masses are given in m$_e$. }
\label{tab:al-line-energy-eff-mass}
\begin{tabular}{l c c}
\hline 
Band & Energy at $\Gamma$ & Effective mass at $\Gamma$  \\

\hline 
1$\Gamma$ & 468  & 2.13 \\
2$\Gamma$ & 196 & 0.40 \\
3$\Gamma$ & 133 &  4.85 \\
E$_{\rm F}$ & 77 &  \\
E$_{\rm CB}$ & 620 &  \\
\hline
\end{tabular}
\end{table}

\end{document}